\documentclass[letterpaper, 10 pt, conference]{ieeeconf}  

\IEEEoverridecommandlockouts                              

\title{\LARGE \bf
Real-time Surgical Environment Enhancement for Robot-Assisted Minimally Invasive Surgery Based on  Super-Resolution}

\author{Ruoxi Wang$^{1}$, Dandan Zhang$^{1}$,~\IEEEmembership{Student Member,~IEEE},  Qingbiao Li$^{2}$, Xiao-Yun Zhou$^{3}$, \\ Benny Lo$^{1}$,~\IEEEmembership{Senior Member,~IEEE}

\thanks{$^{1}$The Hamlyn Centre for Robotic Surgery, Imperial College London, United Kingdom. $^{2}$ University of Cambridge, United Kingdom. $^{3}$ PAII Inc, MA, USA. Corresponding Author email: d.zhang17@imperial.ac.uk. } 
}
\usepackage{graphicx} 
\graphicspath{ {Figures/} }

\usepackage[utf8]{inputenc}
\usepackage[T1]{fontenc}
\usepackage{graphicx}
\usepackage{subfigure}
\usepackage{fancyhdr} 
\usepackage{float}
\usepackage{setspace}
\usepackage{amsmath}
\usepackage{enumerate}
\usepackage{url}
\usepackage{graphicx}
\usepackage{subfigure}
\usepackage{comment}
\usepackage[caption=false,font=footnotesize]{subfig}
 \usepackage{algorithm,algorithmic}
\usepackage{cite}
\usepackage[font=footnotesize]{subfig}
\usepackage{multirow}
\usepackage{ifpdf}
\usepackage{comment}
\usepackage{epstopdf}

\usepackage{algorithmic}
\usepackage{array}
\usepackage{mdwmath}
\usepackage{mdwtab}
\usepackage{eqparbox}
\usepackage{verbatim}

\usepackage{stfloats}
\usepackage[font=footnotesize]{subfig}
\usepackage{comment}
\usepackage{amsmath,bm}
\usepackage{amssymb}
\usepackage{color}
\usepackage{subfig}

\newcommand{\qingbiao}{\textcolor{blue}}

\begin{document}

\maketitle

\begin{abstract}
In Robot-Assisted Minimally Invasive Surgery (RAMIS), a camera assistant is normally required to control the position and zooming ratio of the laparoscope, following the surgeon's instructions. However, moving the laparoscope frequently may lead to unstable and suboptimal views, while the adjustment of zooming ratio may interrupt the workflow of the surgical operation. To this end, we propose a  multi-scale Generative Adversarial Network (GAN)-based video super-resolution method to construct a framework for automatic zooming ratio adjustment. It can provide automatic real-time zooming for high-quality visualization of the Region Of Interest (ROI) during the surgical operation. In the pipeline of the framework, the Kernel Correlation Filter (KCF) tracker is used for tracking the tips of the surgical tools, while the Semi-Global Block Matching (SGBM) based depth estimation and Recurrent Neural Network (RNN)-based context-awareness are developed to determine the upscaling ratio for zooming. The framework is validated with the JIGSAW dataset and Hamlyn Centre Laparoscopic/Endoscopic Video Datasets, with results demonstrating its practicability.


\end{abstract}

\section{Introduction}
In current Robot-Assisted Minimally Invasive Surgery (RAMIS), surgeons have to rely solely from the laparoscopic or endoscopic camera images to carry out the operation. With recent advances in Artificial Intelligence (AI), AI algorithms become increasingly popular in robot-aided MIS surgery to assist surgeons to perform the delicate operations \cite{Zhang2018Self}.

In current practice, surgeons control the surgical robot based on the visual information obtained from the laparoscope. Generally, a camera assistant is required to control and operate the laparoscope following the surgeon's command to focus and zoom into targeted Region of Interest (ROI) during operation. The camera assistant needs to hold the laparoscope and follow the surgeon's instructions to control the position, distance and zooming ratio of the laparoscope, and even for RAMIS, the surgeon has to switch the manipulator to control the camera. However, the zooming ratio required by the surgeons during surgical operations varies, which depends on the specific operation scenarios. A larger zooming ratio is required for delicate operations such as inspecting the tumors or organs, while a smaller ratio might be sufficient for relatively simple operations such as moving the surgical tools to another target area. Indirect visual control is very inefficient for surgical operations. Besides, simply zooming is not enough to provide the surgeon with a  clearer image to assist the operations. Much detailed information, such as the high-frequency fine details, will be lost in the traditional zooming process. In addition, there are only very few laparascopic cameras have optical zooming functions, which can be very costly. Therefore, it is essential to investigate super-resolution methods that could provide the zooming function with high-quality images.

\begin{figure}[ht]
	\centering
	\includegraphics[width = 0.9\hsize]{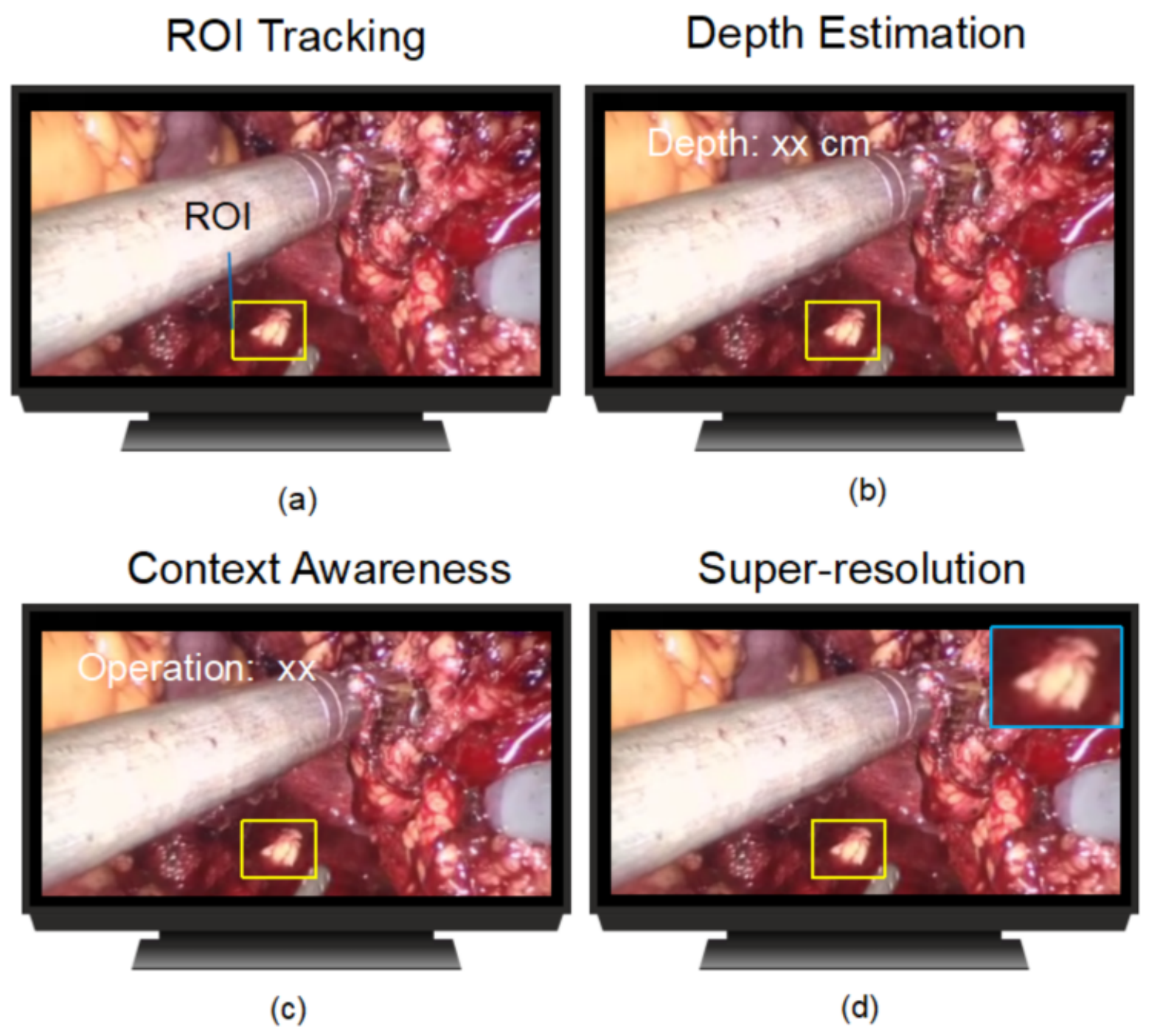}
	\vspace{-1.0em}
	\caption{The pipeline and the typical scenario of the proposed framework. (a) An ROI is initialized and tracked. (b) The depth between the camera and the ROI (xx cm) is estimated. The ROI will not be super-resolved if the estimated depth is less than a pre-defined threshold. (c) The surgeon's next operation will be predicted by RNN-based surgical scene recognition method. The predicted operation will determine the upscaling factor for the super-resolution method. (d) The ROI is super-resolved at a specific upscaling factor determined by the predicted operation and is presented in the top-right corner of the screen or another window. }
	\label{pipeline}
	\vspace{-1em}
\end{figure}


\textbf{Related Work.} End-to-end deep convolutional neural network-based methods have been actively investigated in recent years to solve the super-resolution problem~\cite{P11-8, SRGAN, res3, res4, res5}, of which SRCNN~\cite{SRCNN} and FSRCNN~\cite{FSRCNN} are the most popular methods.

In order to generate more plausible-looking natural images with high perceptual quality, generative adversarial network (GAN) has attracted huge attention in tackling the super-resolution task~\cite{SRGAN, ESRGAN, li2019estimation}. SRGAN~\cite{SRGAN} uses the SRResNet to construct the generator, and uses the content loss and the adversarial loss as the perceptual loss to replace the Mean-Square-Error (MSE) loss to train the generator. To further improve the visual quality of the high-resolution estimation, ESRGAN~\cite{ESRGAN} has been proposed to mainly focus on constructing the generator by using the Residual-in-Residual Dense Block, and the relativistic discriminator to calculate the adversarial loss. 

Some classical video super-resolution approaches are usually considered as optimization methods, which are computationally expensive \cite{FRVSR-2}\cite{FRVSR-11}. The key to solving video super-resolution is to extract the temporal information contained in the adjacent frames. The optical flow is widely used to extract the motion (temporal) information between adjacent frames and warp it onto the current frame, such as the method proposed by Kappeler et al. \cite{FRVSR-20}. Caballero et al. \cite{FRVSR-3} replaced the optical flow method with a trainable motion compensation network, and Mehdi et al. \cite{FRVSR} developed FNet \cite{FNet} to compute the optical flow. However, the methods will be limited to offline data processing due to the network complexity. 

\textbf{Contributions.} In this paper, our primary motivation is to develop an integrated surgical environment enhancement framework, which can assist the surgical procedure.  We focus on developing a real-time multi-scale video super-resolution method, and combining with object tracking, depth estimation and context-awareness method to build a complete framework.  Instead of relying on human interaction, our framework can utilize surgeons' intentions to predict ROI and provide a high-quality, super-resolution view of it automatically. 
In summary, the main contributions of this paper include:

\begin{itemize}
	
	\item Propose a GAN-based real-time multi-scale video super-resolution method;
	
	\item Solve the checkerboard pattern of artifacts in deep-learning-based super-resolution approaches;
	
	\item Build a framework to estimate the ROI and provide high-quality zoom in the image automatically. 
		
\end{itemize}


\begin{figure*}[!ht]
	\centering
	\includegraphics[width = 1\hsize]{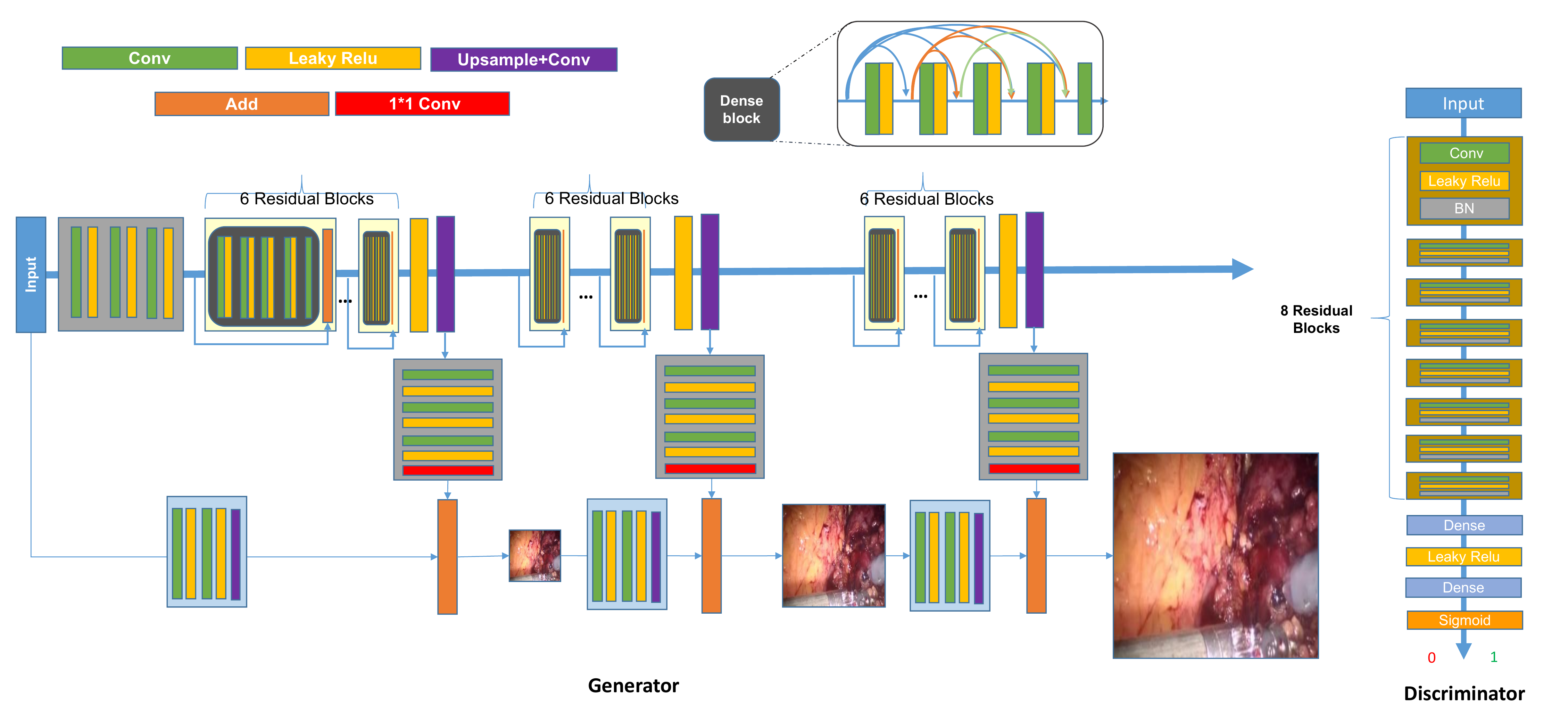}
    \vspace{-2.5em}
	\caption{The architecture of the generator and discriminator network, where the generator consist of a cascaded structure with three different scales.}
	\label{G_structure}
	\vspace{-1em}
\end{figure*}


The remainder of this paper is structured as follows. The methodology and its implementation of the proposed method is described in Sec.~\ref{sec:methodology}. Experiments and results analysis are presented in Sec.~\ref{sec:exp_results}, while conclusions and future works are drawn in Sec.~\ref{sec:Conclusions}.



\section{Methodology}\label{sec:methodology}
%
%
As shown in Fig. \ref{pipeline}, the proposed framework can provide automatic visual control of the camera, which reduces the need of a camera assistant in MIS or the need of surgeon to switch to control the camera in RAMIS.


First, Fig.~\ref{pipeline}.(a) illustrates the tracking of an ROI around the tip of the surgical instrument or the end-effector of the robotic instrument on an initial frame, and Kernel Correlation Filter (KCF) \cite{KCF} is applied to track the instrument tip, which is accurate and fast. KCF uses the cyclic matrix of the area around the target to collect positive and negative samples, and uses ridge regression to train the target detector. The user initially sets the target, then the padding-box around the target is obtained by the cyclic shift, and the one with the greatest response is believed to contain the target. 

Second, as shown in Fig.~\ref{pipeline}.(b), the depth between the camera and the target area is estimated. Camera calibration is performed to rectify the stereo image pairs, then Semi-Global Block Matching (SGBM) \cite{SGBM} is used to compute the disparity map, through which the depth map is obtained by re-projecting the image points to the 3D world. 

Third, as shown in Fig.~\ref{pipeline}.(c), surgical scene recognition and the intentional surgical movements are predicted with a Recurrent Neural Network (RNN) based module. As RNN is designed to handle sequential data, it can take a sequence as input, and extracts the spatial and temporal features to model the mapping for surgical gesture recognition and prediction.


Finally, the depth value and the surgeme recognition results are used to determine the upscaling ratio of the ROI for the automatic zooming. To solve the image blur caused by zooming, a multi-scale GAN-based video super-resolution method is developed to enhance the visual quality of generated images, as demonstrated in Fig.~\ref{pipeline}.(d) and described as follows.


\subsection{Adversarial network architecture}\label{sec:network_archi}

 \subsubsection{Generator}
As shown in Fig. \ref{G_structure}, a generator $G$ with a multi-scale structure is constructed based on the Laplacian pyramid to generate super-resolved images of different resolutions. It consists of $n$ sub-networks for super-resolving a low-resolution image with an upscaling factor of $2^n, n\in[1, N]$, where $N$ is the number of cascade level. For example, when $N = 3$, the upscaling factor contains 2, 4, and 8. There are two branches in the network: the top one is for feature extraction, while the bottom one is for image reconstruction. In the feature extraction phase, the residual block layout introduced in \cite{SRGAN-24} is used to construct the $M$ residual blocks. 

The dense block \cite{huang2017densely} is used to enhance the feature transfer efficiency between layers, consisting of convolutional layers with $3\times3$ kernels and $64$ feature channels followed by a \texttt{LeakyReLu} activation function. The dense connection enables the inputs of each layer to be derived from the output of all previous layers. Each layer directly accesses the gradients from the loss function and the original input signal, leading to implicit deep supervision and better performance. 

The \texttt{up-sample+convolution} block is used for up-sample the extracted features to further predict residual features at level $n$ and extract more in-depth features at level $n+1$. The cascade structure increases the non-linearity of $G$ to learn more complex mappings for higher levels. In the image generation phase, the input is up-sampled to higher resolution and RGB channels and then combined (element-wise sum) with the residual image obtained from the feature extraction phase to produce the super-resolved image $I_s^{SR}$ at level $s$. Finally, $I_s^{SR}$ is fed into $s+1$ level to generate $I_{s+1}^{SR}$.

\begin{figure}[ht]
    \centering
    \includegraphics[width = 1.0\hsize]{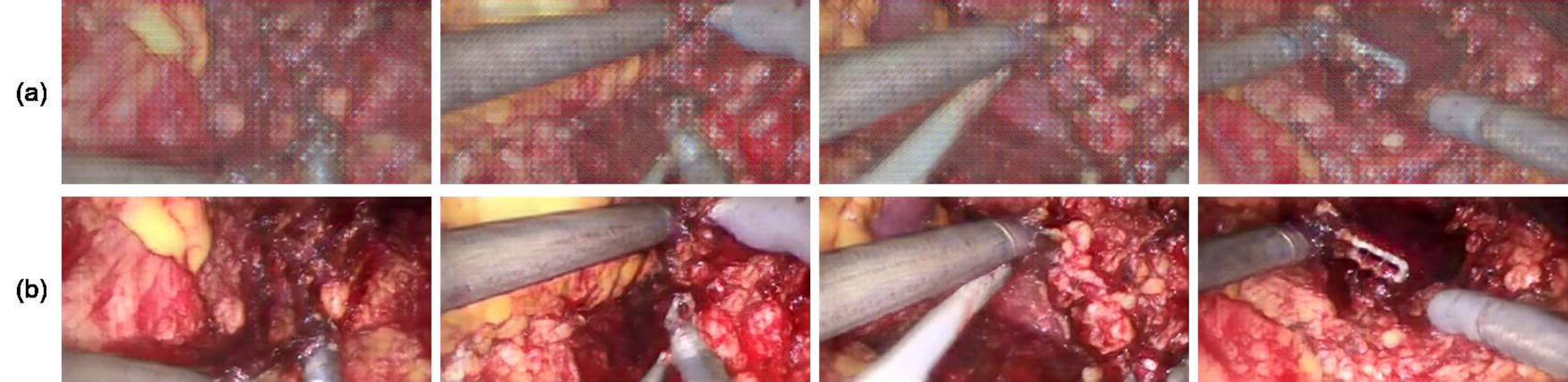}
      \vspace{-2em}
    \caption{(a): The checkerboard pattern of artifacts occurs in the generated images by deconvolution. (b):  The checkerboard artifacts are eliminate in the generated image by \texttt{upsample+convolution}. }
    \label{artifact}
    \vspace{-2em}
\end{figure}

In the experiments, a checkerboard pattern of the artifact is observed in the images that were produced by the generator $G$, as presented in Fig. \ref{artifact}.(a). It is more obvious in surgical images because the checkerboard pattern tends to be more prominent in images with strong colors.

The process of deconvolution may lead to this artifact. \texttt{PixelShuffle} \cite{ESPCN}
(sub-pixel convolution) is used in ESRGAN, and \texttt{Conv2DTranspose} \cite{DCGAN} is used in LapSRN, both of which contain deconvolution operation. However, deconvolution can easily produce uneven overlap, which puts more paint at some points than others \cite{deconv}. Particularly, deconvolution will cause uneven overlap when the kernel size is not divisible by the stride. This artifact is still likely to occur, even the deconvolution with divisible kernel size and stride. Hence, we carried out an experiment where the deconvolution is replaced by \texttt{upsample+convolution} to verify the above conjecture. Fig. \ref{artifact} (b) demonstrated the image generated by the generator without deconvolution. Although, in principle, the network using deconvolution could carefully learn the weights to avoid this artifact, it will scarify the model capacity and is hard to achieve. Therefore, the \texttt{upsample + convolution} is used to replace deconvolution. 


\subsubsection{Discriminator}
Unlike the standard discriminator that identifies an input image is real or fake, the relativistic discriminator $D(.)$ is used here to estimate the probability that a real image $I_r$ is more realistic than a fake image $I_f$. It can be defined as: 
\begin{equation}\label{eq:discriminator_false}
    D_{Ra}(I_r , I_f) = \sigma (C(I_r) - E[C(I_f)]) \rightarrow 1
\end{equation}
\begin{equation}\label{eq:discriminator_true}
    D_{Ra}(I_f , I_r) = \sigma (C(I_f) - E[C(I_r)]) \rightarrow 0 
\end{equation}
where $\sigma$ is the sigmoid function and $C(.)$ is the non-transformed discriminator output. Eq.~\ref{eq:discriminator_false} indicates that the real data is more realistic than fake data, and Eq.~\ref{eq:discriminator_true} implies that fake data is less realistic than the real data. The countereffect between generator and discriminator encouraged discriminator to estimate that the real image is more realistic than the fake one, and therefore, the generator is learned to produce images with higher quality and photo-realistic.
\subsection{Perceptual loss function}\label{sec:loss_fun}

Perceptual loss \cite{ESRGAN-13} is widely used in the SR problem, because only using pixel-wise loss function (i.e., MSE) can cause a lack of fine high-frequency details in reconstructed images. 
The perceptual loss function is introduced as a weighted sum of three different loss functions here, including pixel loss, content loss, and adversarial loss. It can be defined as follows:
\begin{equation} \label{eq:loss_sr}
    L^{SR} = \mu L_{pixel} + \eta  L_{content} + \lambda  L_{adversarial}
\end{equation}
where the coefficients of the three components ($\mu$, $\eta$ and $\lambda$) are hyper-parameters and are tuned empirically. 
\subsubsection{Pixel loss}
The pixel loss is defined as the pixel-wise loss between the real images and the generated images:
\begin{equation}
    L_{pixel} = \frac{1}{r^2WH} \sum_{x=1}^{rW} \sum_{y=1}^{rH} \rho(I_{x,y}^{HR}, G(I^{LR})_{x,y})
\end{equation}
where $W$ and $H$ represent the width and height of the high-resolution image, and $r$ represents the scaling factor. $\rho$ is the L1 Charbonnier loss function.

\subsubsection{Content loss}

A deeper pre-trained VGG19 layer is employed for a deeper image generation level in the generator network. For instance, the $7^{th}$ layer of VGG19 is used for level 2 (4x upscaling output), and the $16^{th}$ layer of VGG19 is used for level 3 (8x upscaling output). Therefore, the content loss is defined in Eq.~\ref{eq:loss_content} based on the convolutional layers of the pre-trained VGG19 network, as follows.
\begin{equation} \label{eq:loss_content}
\begin{split}
    L_{\text{content}} = \frac{1}{W_{i}H_{i}} \sum_{x=1}^{W_{i}} \sum_{y=1}^{H_{i}}\rho(\varphi_{i}(I^{HR})_{x,y}, ~
    \varphi_{i}(G(I^{LR}))_{x,y})
\end{split}
\end{equation}
where $\varphi_{i}$ is defined as the feature map obtained by the i-th convolutional layer of the VGG19 network. The content loss is defined as the L1 Charbonnier loss $\rho$ between the real image $I^{HR}$ and the generated image $G(I^{LR})$. $W_i$ and $H_i$ indicates the dimensions of the corresponding feature map in the VGG19 network. 

\begin{figure*}[ht]
\centering
\includegraphics[width = \textwidth]{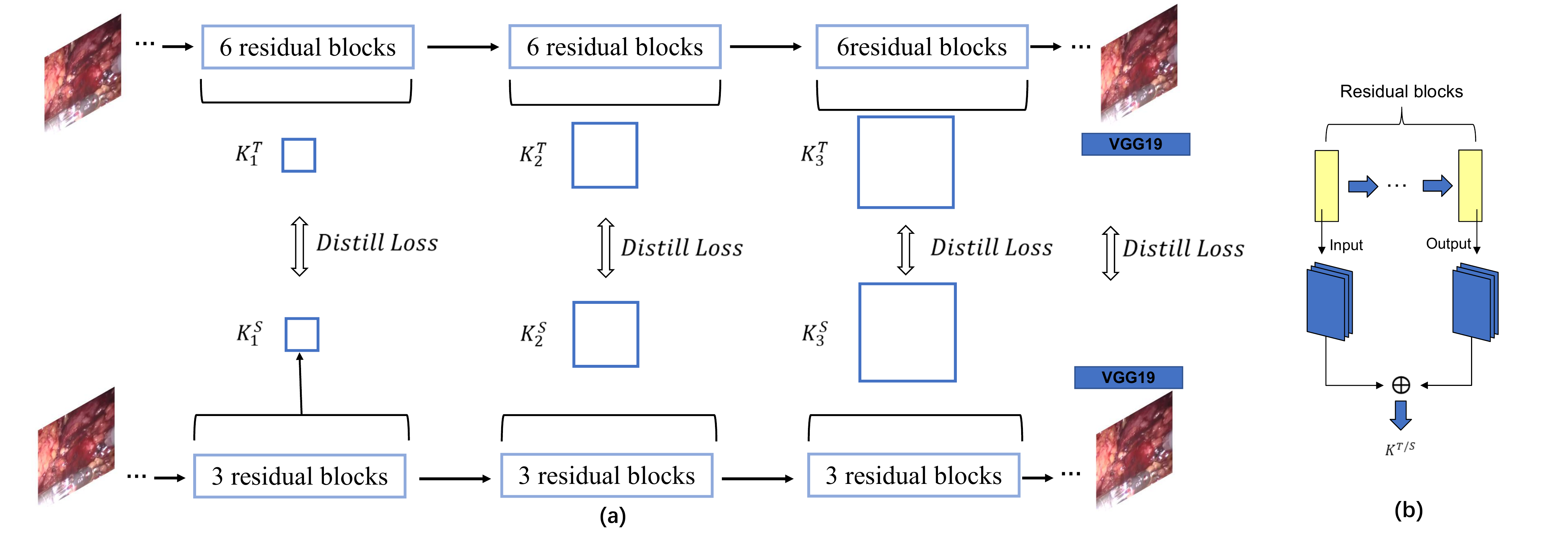}
\vspace{-2.5em}
\caption{ (a) The strategy of the proposed knowledge distillation method. L2 loss is used to calculate the distill loss. (b) The illustration of the proposed method to distill knowledge from intermediate layers, where the number of residual modules can be changed.}
\vspace{-2em}
\label{KD}

\end{figure*}

\subsubsection{Adversarial loss}
The adversarial loss is defined by the adversarial training between $G$ and $D(.)$.
\begin{equation}
\begin{split}
    L_{\text{adversarial}} = -\mathbb{E}_{x_r}[log(1-D_{Ra}(x_r, x_f))] -\\
    \mathbb{E}_{x_f}[log(D_{Ra}(x_f,x_r))]
\end{split}
\end{equation}
where $D_{Ra}$ is the relativistic discriminator loss function. Real image $x_r$ and generated image $x_f$ are all involved in the adversarial loss for the generator, thereby enabling the generator to be optimized from the gradients from both real and generated data. 

For the pixel and content loss, the robust L1 Charbonnier loss is calculated, since it could better handle outliers and improve the super-resolution performance over the L2 or L1 loss function \cite{L1charbonnier}. 
\begin{equation}
\begin{split}
L_{1_{\text{Charbonnier\, Loss}}} &= \rho (y_{true},y_{pred})\\& = 
\sqrt{(y_{true}-y_{pred})^2 + \epsilon ^2}
\end{split}
\end{equation}
where $\epsilon$ is a small constant that is used to stabilize the result. 
%
%
\vspace{-1em}
\subsection{Video super-resolution}\label{sec:Video_super_resolution}



\begin{figure}[h]
	\centering
	\includegraphics[width = 0.85\columnwidth]{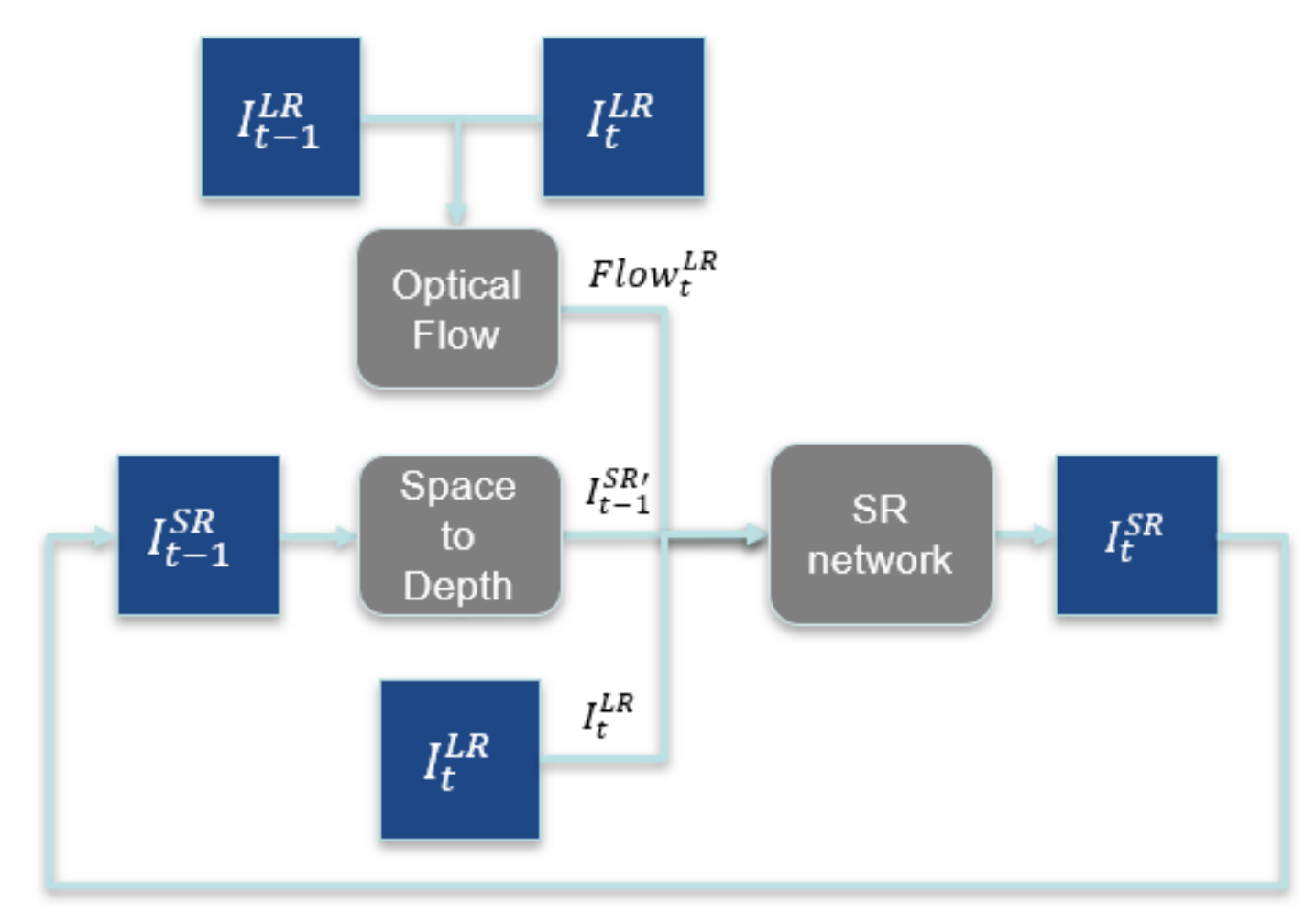}
	\vspace{-1em}
	\caption{Framework of the video super-resolution. }
	\label{video SR}
	\vspace{-2em}
\end{figure}

Instead of producing individual images, an online recurrent video super-resolution framework is proposed, which utilizes the optical flow to recover more high-frequency details by considering the temporal relation among different video frames. Different from the approaches introduced by Caballero \textit{et al.} \cite{FRVSR-3} and Mehdi \textit{et al.} \cite{FRVSR}, conventional optical flow method is deployed to improve the efficiency of the framework. Hence, our super-resolution GAN is the only trainable component.
Fig.~\ref{video SR} illustrated our pipeline for video super resolution, which are also summarised as follows. 

Firstly, the optical flow~\cite{farneback2003two}, $\mathrm{Flow}^{LR}_{t}$, is computed by estimating the motion between low-resolution input $I_{t-1}^{LR}$ and $I_{t}^{LR}$. Here, the flow can represent the displacement of each pixel in $x$ and $y$ directions. 

Secondly,  to improve the efficiency of the model, the super-resolved high-resolution images ($I_{t-1}^{SR}$) at previous time frame is mapped to low-resolution space into $I_{t-1}^{SR'}$ using space-to-depth transformation by $rH \times rW \times C \rightarrow H \times W \times r^2C$. 

Finally,  the low-resolution from the previous super-resolved image ($I_{t-1}^{SR'}$), the estimated flow ($\mathrm{Flow}_{t}^{LR}$), and the current low-resolution frame ($I_{t}^{LR}$) are concatenated as $64\times64\times53$. Then, the concatenated image is fed into our super-resolution network to generate the high-resolution estimation of the current frame $I_{t}^{SR}$.  
Note that $Flow^{LR}$ and $I_{t-1}^{SR'}$ are initialized with zeros at the first time frame.

\subsection{Knowledge distillation} \label{sec:knowledge_distillation}


To further improve the speed of executing the proposed algorithm, a knowledge distillation method  \cite{KD} (Fig. \ref{KD}.(a)) is introduced to compress the model while preserving its performance. Generally, transferring knowledge is to let the student model mimic the intermediate results of the teacher model. 
The knowledge distillation process consist of two stages: in the first stage, the distilled knowledge is extracted from the teacher model and the student model using the L2 loss as the distill loss to train the student model. In the second stage, the pre-trained student model will be trained again with the retained weights. 

In our framework, the residual blocks in the generator network are divided into three sections depending on their spatial dimensions. As demonstrated in Fig. \ref{KD}.(b), the distilled knowledge is defined as the features extracted from the first and last layers of each section so that the student model can learn the intermediate results and the process of the solution.
Besides, the pre-trained VGG19 network is also used to extract the features of the final output from the student and teacher model to calculate the distill loss.

\section{Experiments and Results}\label{sec:exp_results}
\subsection{Dataset}


For training the super-resolution model, real surgical images are used as our dataset. A dataset \cite{hamlyn-dataset} with 40,000 pairs of rectified stereo images are collected during the partial nephrectomy with the da Vinci surgical system by the Hamlyn Centre (Imperial College London, UK), which is used to train the super-resolution model. The original resolution of images is $384 \times 192$. Specifically, 7000 surgical images are used as the training set, and 3000 surgical images are used as the evaluation and testing set. For quantitative analysis, the results are evaluated with two widely used image quality metrics: Peak Signal-to-Noise Ratio (PSNR)~\cite{PSNR} and Structural Similarity Index Measure (SSIM)~\cite{SSIM}.


\subsection{Implementation}

The network is trained on an NVIDIA Tesla P100-PCI-E GPU with TensorFlow 1.14.0, CUDA 10.1. A three-level cascade structure is constructed for the generator $G$ to produce $2\times$, $4\times$, and $8\times$ upscaling super-resolved high-resolution results, where the dimensions of the input image is $64 \times 64 \times 3$. The low-resolution input images are obtained by down-sampling the high-resolution images using bilinear interpolation with down-sampling factors 2, 4, and 8, respectively.  Note that, the discriminator $D(.)$ is only trained by the $8\times$ super-resolution result. The $7^{th}$ layer of VGG19 is used to calculate the content loss for the $2\times$ and $4\times$ super-resolution results, and the $16^{th}$ layer for the 8x super-resolution result. The coefficients of the perceptual loss in Eq.~\ref{eq:loss_sr} in is defined as $\mu = 0.1$, $\eta=1$ and $\lambda=0.001$, respectively. 
As mentioned in Sec.~\ref{sec:knowledge_distillation}, the optical flow, and the previous super-resolved high-resolution image are initialized with zeros to estimate the video at the first frame $I_1^{SR}$.  After obtaining a deep model, the proposed knowledge distillation method is used to train a relatively light model to increase the inference speed. 
\vspace{-0.5em}

\subsection{Evaluation of multi-scale super-solution model}\label{sec:results_qual_quant}
\begin{table}[ht]
    \vspace{-1em}
    \centering
    \caption{The comparison of the proposed method and several SISR methods on the BSD200 dataset.}
     \vspace{-1em}
    \scalebox{0.75}{
    \begin{tabular}{|c|c|c|c|c|c|c|}
        \hline
         BSD200 & Nearest & Bicubic & SRCNN & SRResNet & SRGAN & \textbf{Proposed}  \\
          \hline
         PSNR & 25.02 & 25.94 & 26.68 & 27.58 & 25.16 & \textbf{28.22}  \\
         \hline
         SSIM & 0.6606 & 0.6935 & 0.7291 & 0.7620 & 0.6688 & \textbf{0.8902}\\
         \hline
    \end{tabular}}
    \vspace{-1em}
    \label{SISR_compare}
\end{table}

\begin{table}[ht]
    \vspace{-1em}
    \centering
    \caption{Quantitative evaluation of our network at multiple scales}
    \vspace{-1em}
    \scalebox{0.75}{\begin{tabular}{|c|c|c|c|}
        \hline
          & 2X & 4X & 8X \\
          \hline
         PSNR & 30.02 & 31.86 & 32.72  \\
         \hline
         SSIM & 0.9293 & 0.9407 & 0.9532\\
         \hline
    \end{tabular}}
    \vspace{-1em}
    \label{psnr_ssim}
\end{table}

\begin{figure}[ht]
	\centering
	\includegraphics[width = 1\hsize]{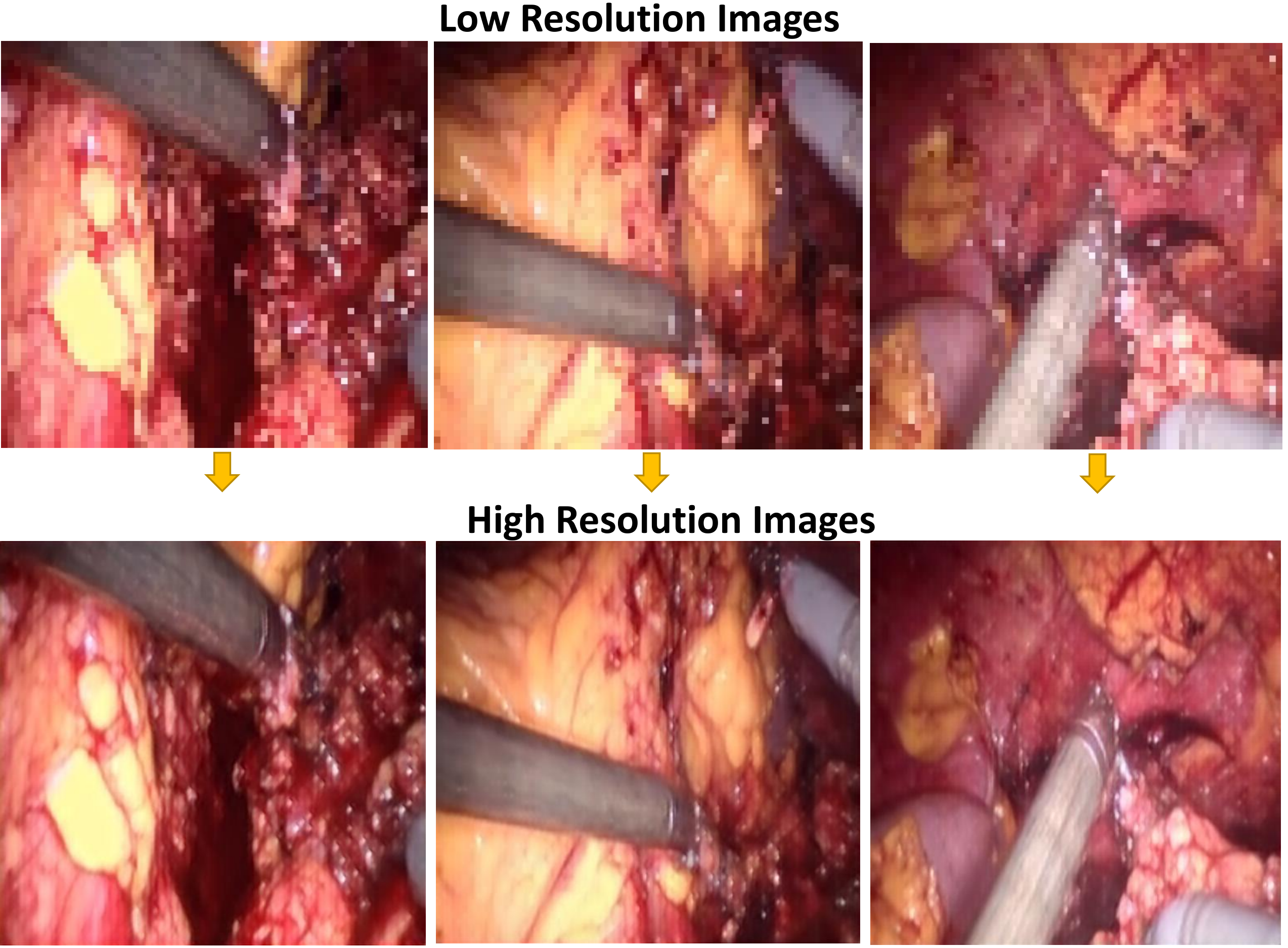}
	\vspace{-2em}
	\caption{Demonstration for the translation from low resolution images to high resolution ones.}
	\label{quantitative results}
	\vspace{-2em}
\end{figure}



To evaluate the performance of the proposed pipeline, the video super-resolution framework in Sec.~\ref{sec:Video_super_resolution} is replaced by various Single-Image Super-Resolution (SISR) approach on the BSD200 dataset. Table~\ref{SISR_compare} summaries the comparison between the performance of our framework and that of other SISR approaches, including the nearest,  bicubic interpolation, SRCNN \cite{SRCNN}, SRResNet and SRGAN \cite{SRGAN}.
PSNR \cite{PSNR} is a widely used image objective evaluation index, which mainly estimate the pixel-level error.
Considering the visual characteristics of human eyes, we also used SSIM \cite{SSIM} to quantitatively measure image similarity in terms of brightness, contrast, and structure.
The PSNR and SSIM of the proposed method are $28.22$ and $0.8902$, which are higher than all the other results, thereby proving that the proposed method has better performance than the other approaches.


Therefore, our video super-resolution method can be used to super-resolved surgical images with different upscaling factors. The inference speed reaches 28 fps. For quantitative evaluation, Table~\ref{psnr_ssim} shows the statistics about PSNR and SSIM of the super-resolved images at $2\times$, $4\times$, and $8\times$ upscaling level using our method.
Fig. \ref{quantitative results} presented some examples of generated images at multi-scale, where the example of a super-resolved video can be accessed via \url{https://youtu.be/VwIwVwTLteE}.

\begin{figure*}[ht]
	\centering
	\includegraphics[width = 1\hsize]{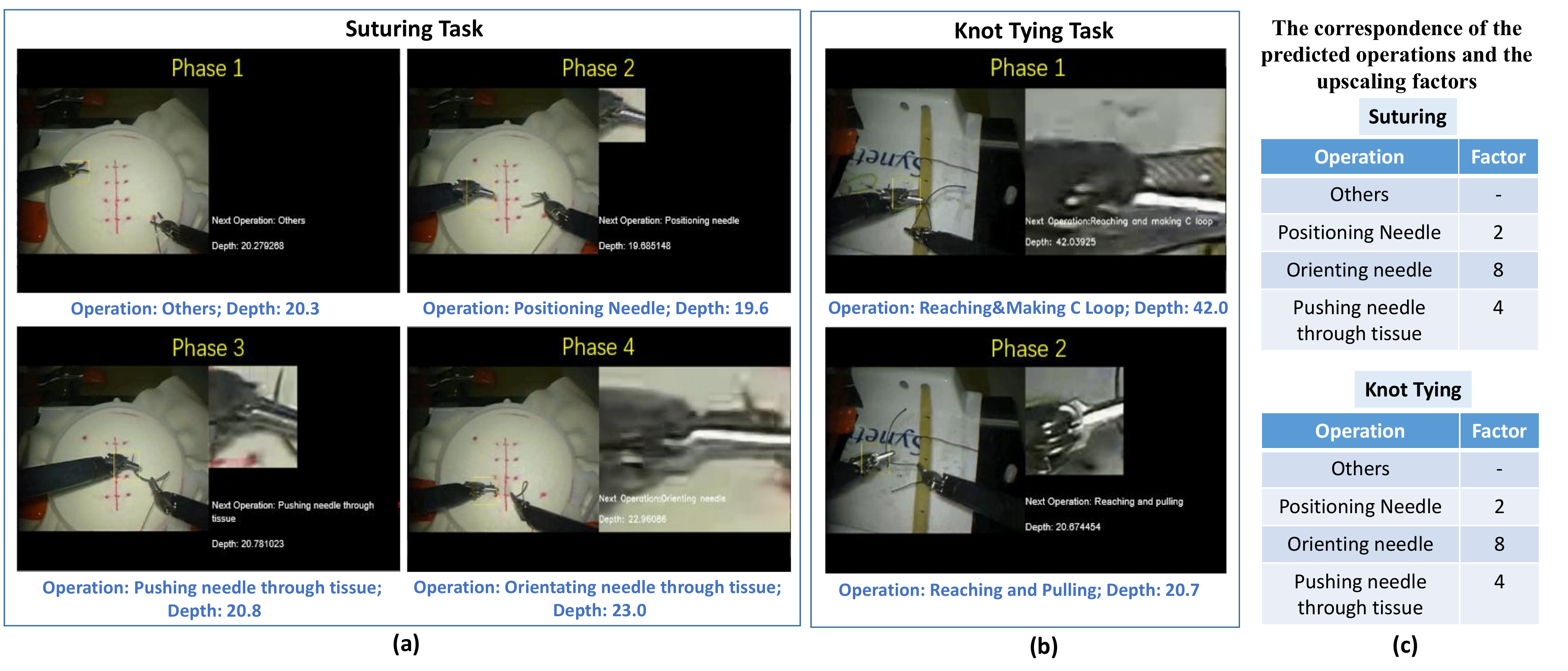}
	\vspace{-2.5em}
	\caption{Experiments of our pipeline in JIGASW dataset, demonstrating different phases of testing our pipeline in the suturing (a) and knot tying (b) tasks, while (c) summaries the correspondence of the predicted operations and the upscaling factors.}
	\label{phases.png}
	\vspace{-2em}
\end{figure*}

\vspace{-0.5em}
\subsection{Evaluation of the overall workflow}\label{sec:result_full_workflow}

We test our framework on the JIGSAW dataset \cite{JIGSAW}, which contains video and kinematic data of four different surgical tasks. In this dataset, the annotations are provided to describe the ground truth gesture (operation) at each frame. We also train our super-resolution model on this video dataset, and the demo of its performance can be accessed via \url{https://youtu.be/UChxJVeWd _ s}.

This video demonstrated the whole workflow of our pipeline, as we proposed in Fig.~\ref{pipeline}. Firstly, an ROI is initialized and tracked by the KCF tracker. The estimated depth and the predicted surgemes are written at the bottom of the screen. Suppose the depth is greater than the pre-defined threshold, which we set as $10\,cm$. In that case, the low-resolution ROI will be super-resolved by the super-resolution method with the upscaling factor depending on the predicted operation. The correspondence of the operation and the upscaling factor are summarized in Fig. \ref{phases.png}(c). In the meanwhile, the super-resolved image will be shown in another window. Our framework can achieve real-time in the inference stage in our experiments with an inference speed of around 25 fps.

Fig. \ref{phases.png} demonstrated different phases when inferencing our framework in the dataset. 
Fig. \ref{phases.png} (a) shows the four phases in the suturing task. In Phase 1 (a), an ROI is initialized and located by a yellow box. The estimated depth is $20.27\,cm$, and the predicted next operation is \textbf{others}. In this case, there is no need to super-resolve the ROI according to Fig. \ref{phases.png} (c).  In Phase 2 (Fig.~\ref{phases.png}.(a)), the predicted operation based on RNN is \textbf{Positioning needle}, and the estimated depth is $19.6\,cm$. In this case, the ROI is super-resolved by the upscaling factor of 2. Following that, surgeons need to push the needle through the tissue and orient the needle from the inside. Our framework can correctly predict \textbf{Pushing needle through tissue} in phase 3 of Fig.~\ref{phases.png}.(a) with estimated depth $20.8\,cm$, and \textbf{Orienting needle} in phase 4 of Fig.~\ref{phases.png}.(a) with estimated depth $23.0\,cm$, respectively. As illustrated in Fig.~\ref{phases.png}.(a), our model set the upscaling factor as $4$ for \textbf{Pushing needle through tissue} in phase 3 and the upscaling factor as $8$ for \textbf{Orienting needle} in phase 4. 

Besides, our framework can be used in other surgical tasks, such as knot-tying tasks. Fig. \ref{phases.png} (b) demonstrates that appropriate zoom-in scales are chosen based on the predicted action and estimated depth. 
More details of the results can be viewed in the supplementary videos.

\subsection{Discussions}
The effectiveness of the proposed framework has been verified qualitatively and quantitatively. However, we observe that the quality of the super-resolved results depends heavily on the training and test data quality. High-quality ground truth is essential to train the model so that it can obtain superior perceptual performance. When the test data has a very low resolution or is significantly different from the training set, the super-resolution algorithm may fail in producing the desired results. In addition to improving the super-resolution results, the effectiveness of the proposed framework can also be influenced by tracking accuracy of ROI, the depth estimation, and the context-awareness accuracy. As surgical scenarios are getting increasingly complex, simply tracking one or more target areas is not practical. Gaze tracking might be a potentially effective way to identify the ROI, since surgeons normally focus on the target area for operation during the surgical operation. Advanced algorithms for precise depth estimation and context-awareness can be incorporated into the framework, while the correspondence of the predicted operation and the upscaling factors, as well as the threshold of the depth value and, can be customized by surgeons during clinical applications. Other AI techniques can be incorporated into the framework to assist the development of intelligent robotic assistants \cite{Zhang2020Automatic}.



\section{Conclusions}\label{sec:Conclusions}
\vspace{-0.5em}






In this paper,  we present a framework for automatically super-resolving the ROI with suitable upscaling factors to improve surgical efficiency. A multi-scale video super-resolution method is proposed, which can produce high-quality images in real-time without the common checkerboard artifacts that occur in other super-resolution methods~\cite{ESPCN,DCGAN,deconv}. 
The standard quantitative and qualitative evaluations in Sec.~\ref{sec:results_qual_quant} have verified the superior performance of our proposed super-resolution method against other SISR approaches. Preliminary experiments show that deeper networks can produce higher visual quality images, while the network complexity restricts its inference speed. In the experiments, we find a balance between the speed and the performance by applying the knowledge distillation method. 
The overall statistics and video demo in Sec.~\ref{sec:result_full_workflow} have proven that the framework is feasible, effective, and practical for enhancing the surgical environment for RAMIS. 
  
Future work will include collecting more data with higher quality to train the super-resolution model and adapting the proposed framework to various surgical robotic systems for further validation. Surgeons will be invited to participate in the user studies for potential clinical applications in different types of surgery.

\bibliographystyle{IEEEtran}
\bibliography{IEEEabrv,ref}

\end{document}